\pdfoutput=1
\documentclass[12pt, preprint]{aastex}
\shorttitle{Radio imaging of a hot-channel structure}
\shortauthors{Vasanth et al.}

\usepackage{graphicx}
\usepackage{color}
\usepackage[normalem]{ulem}

\begin{document}


\title{An Eruptive Hot-Channel Structure Observed at Metric Wavelength as a Moving Type-IV Solar Radio Burst}


\author{V. Vasanth\altaffilmark{1}, Yao Chen\altaffilmark{1} $^{\dagger}$, Shiwei Feng\altaffilmark{1}, Suli Ma\altaffilmark{2},
Guohui Du\altaffilmark{1}, Hongqiang Song\altaffilmark{1},
Xiangliang Kong\altaffilmark{1}, and Bing Wang\altaffilmark{1}}

\altaffiltext{1}{Institute of Space Sciences, Shandong University,
Weihai, Shandong 264209, China yaochen@sdu.edu.cn,
vasanth@sdu.edu.cn}

\altaffiltext{2}{College of Science, China University of
Petroleum, Qingdao 266580, China}

\begin{abstract}

Hot channel (HC) structure, observed in the high-temperature
passbands of the AIA/SDO, is regarded as one candidate of coronal
flux rope which is an essential element of solar eruptions. Here
we present the first radio imaging study of an HC structure in the
metric wavelength. The associated radio emission manifests as a
moving type-IV (t-IVm) burst. We show that the radio sources
co-move outwards with the HC, indicating that the t-IV emitting
energetic electrons are efficiently trapped within the structure.
The t-IV sources at different frequencies present no considerable
spatial dispersion during the early stage of the event, while the
sources spread gradually along the eruptive HC structure at later
stage with significant spatial dispersion. The t-IV bursts are
characterized by a relatively-high brightness temperature ($\sim$
10$^{7}$ $\--$ 10$^{9}$ K), a moderate polarization,
and a spectral shape that evolves considerably with time. This
study demonstrates the possibility of imaging the eruptive HC
structure at the metric wavelength and provides strong constraints
on the t-IV emision mechanism, which, if understood, can be used
to diagnose the essential parameters of the eruptive structure.
\end{abstract}

\keywords{Sun: corona --- Sun: activity --- Sun: coronal mass
ejections (CMEs) --- Sun: radio radiation}

\section{Introduction}

Hot channel (HC) structures, first observed through the
high-temperature passbands at 131 and 94~\,\AA{} of the
Atmospheric Imaging Assembly (AIA: Lemen et al. 2012) on the Solar
Dynamic Observatory (SDO: Pesnell et al. 2012) (Cheng et al. 2011;
Zhang et al. 2012), are regarded as one candidate of coronal flux
ropes that are the main energy carrier and one of the essential
elements of solar eruptions. It is not or hardly present in cooler
passbands of AIA such as the 211 and 171 {\AA}, indicating its
temperature is as high as 5 $\--$ 10 MK. The formation process and
the associated heating mechanism of HCs, not resolved at this
time, are likely related to some slow reconnections or volume
current dissipations before or during the early stage of a flare.
It is natural to further ask whether energetic electrons are also
generated during the HC formation/heating and eruptive process and
whether these energetic electrons can excite radio emissions. If
yes, this may provide an additional observational window of the
eruptive HC-flux rope structure, important to further diagnose its
physical properties and explore its formation and eruptive
mechanism, as well as the distribution of energetic electrons
within the structure.

A latest study by Wu et al. (2016) shows that during the
pre-impulsive stage the HC structure can be imaged in the
microwave regime using the Nobeyama Radioheliograph (NoRH;
Nakajima et al. 1994; Takano et al. 1997). They found that the HC
at 17 GHz presents an overall-arcade like structure with several
intensity enhancements bridged by relatively weak emission. Based
on their analysis, the microwave emission within the HC is at
least partly contributed to by non-thermal energetic electrons.
The study thus presents a piece of observational evidence for the
existence of non-thermal energetic electrons within the HCs. Here
we continue to expand the radio imaging study on HC structures to
the metric wavelength that presents a direct observational
manifestation of energetic electrons in the inner corona.

It is well known that there exist five types of radio bursts at
the metric wavelength, including type-I, II, III, IV, and V bursts
(Wild {\&} McCready 1950; Melrose 1980; McLean {\&} Labrum, 1985).
Among them, the type IV (t-IV) radio burst is the most likely type
related to the eruptive flux rope structure, while the type I
burst is related to active regions (ARs) often in the absence of
strong solar activities, the type II burst is attributed to
energetic electrons accelerated at coronal shocks (see Feng et al.
2013, 2015; Kong et al. 2012; Chen et al. 2014; Vasanth et al.
2014; Du et al. 2015 for latest studies), and type III and V
bursts are given by energetic electrons accelerated at solar
flares propagating outward along open field lines. The t-IV bursts
are broadband continuum emission, further classified as the moving
t-IV (t-IVm) and static t-IV bursts (Weiss 1963; Melrose, 1980),
generally believed to be excited by energetic electrons trapped
within certain magnetic structures, such as magnetic arches,
loops, or plasmoid structures (Smerd {\&} Dulk 1971; Vlahos et al.
1982; Stewart 1985). Different emitting mechanisms have been
proposed, including the synchrotron and gyro-synchrotron emission
(Boischot 1957; Bastian {\&} Gray 1997; Bastian et al. 2001), the
plasma emission (Duncan 1981; Stewart 1982; Ramesh et al. 2013),
and the electron cyclotron maser emission (Kujipers 1975; Lakhina
{\&} Buti 1985; Winglee {\&} Dulk 1986).

Latest observational studies start to conduct combining analysis
of the imaging data obtained by Nan\c{c}ay Radioheliograph (NRH;
Kerdraon {\&} Delouis 1997) at the metric wavelength (Tun {\&}
Voulidas 2013; Bain et al. 2014) and the high-quality EUV data
given by SDO/AIA. Yet, due to the scarcity of events with a
complete data set, till now only one t-IV event has been
investigated along this line (Tun {\&} Voulidas 2013; Bain et al.
2014). The two studies investigated the same t-IVm event and
concluded that the gyro-cyclotron emission is responsible for the
burst. This is based on the observational characteristics such as
the relatively low brightness temperature ($T_B$ $<$ 10$^{6}$ K),
the absence of spatial dispersion at different NRH imaging
frequencies, and the power-law spectra typical for
gyro-synchrotron emission. The radio sources are found to be
co-spatial with an eruptive filament that may represent the core
of the coronal mass ejection (CME).

Here we show another moving t-IV burst with properties distinct
from those investigated by Tun {\&} Voulidas (2013) and Bain et
al. (2014). The burst is found to be closely associated with an
eruptive HC structure, thus establishing the physical connection
between the t-IV sources and the HC-flux rope, and demonstrating
the possibility of imaging this essential component of solar
eruptions at metric wavelength.

\section{Overview of the Event: the HC eruption and the radio burst}
The eruption originated in NOAA AR11429 at the northeastern limb
on 2012 March 4. See Figure 1 (and the accompanying movie) for the
AIA observed eruptive process at two hot passpands 131 and
94~\,\AA{} and one cool passband 171~\,\AA{}. This AR, well
studied by a number of authors (e.g., Harker {\&} Pevtsov 2013;
Liu et al. 2014; Colaninno {\&} Vourlidas 2015), is super-active
releasing 15 M-class and 3 X-class flares and several CMEs during
its transit across the disk from March 2 to March 17. The eruption
studied here is associated with an M2-class flare located at
N16E65 and a halo CME propagating at an average speed of 1306 km
s$^{-1}$ according to the Large Angle Spectroscopic Coronagraph
(LASCO; Brueckner et al. 1995) C2 data. According to the GOES SXR
light curves (see Figure 2a), the flare started at 10:29 UT and
peaked at 10:52 UT lasting for nearly two hours.

The pre-eruption structure, clearly visible from the 94~\,\AA{}
image in the upper panels of Figure 1 (pointed at by a pair of red
arrows), starts to rise slowly at $\sim$10:27 UT, several minutes
before the flare start. The structure presents a highly curved
morphology writhing around the main bright loops of the
AR. It has one root at the northwestern side of the AR, while the
other root is unidentifiable and possibly located behind the AR.
Along with its slow rise, the structure becomes clearly visible at
131~\,\AA{}, yet remaining invisible at 171~\,\AA{}. This
indicates that the structure contains high-temperature plasmas.

Panels of Figure 1b present the eruptive structure. The
original writhed structure evolves into a bright plasmoid
connected to a co-moving faint arcade feature on its southern side
that is clearly visible from the accompanying movie. The eruptive
structure is best seen at 131~\,\AA{} (and invisible at
171~\,\AA{}), again indicating its high internal temperature.
It represents an eruptive HC structure (Zhang et al.
2012), likely associated with a magnetic flux rope consisting of
twisted field lines with free magnetic energy to power the
subsequent solar eruption.

During the eruption, the HC gradually fades and becomes invisible
around 10:45 UT, while at its northwestern foot there appears an
obvious dimming region (see green arrows in Figure 1c). This
indicates that the dimming is due to an evacuation of plasmas. The
dimming region is surrounded by a narrow layer of brightenings
that evolves dynamically, and expands slowly towards the
northwestern direction during the eruption. The brightening layer
is likely attributed to the dissipation of currents (or magnetic
reconnection) at the interface of the HC and the surrounding
magnetic field.

In Figure 2a, we also present the Fermi Gamma-ray Burst
Monitor (GBM) (Meegan et al. 2009) data at three energy bands.
This is to show the presence of HXR-emitting non-thermal
electrons, as well as the timing of the flare. In Figure 2b-2c,
we show distance-time maps along the slice S1 (see Figure 1b),
which corresponds to the outward moving direction of the plasmoid
of the HC. The overlying arcade that has been pushed outward by
the HC (see black arrows in Figure 1) is clearly visible in the
S1-171~\,\AA{} map. The motion of the plasmoid can be separated
into a slow rise stage at a speed of $\sim$ 72 km s$^{-1}$ and an
eruptive stage at $\sim$ 367 km s$^{-1}$, while the overlying
171~\,\AA{} arcade expands at slightly smaller speeds of $\sim$ 56
km s$^{-1}$ and 335 km s$^{-1}$ (along S1), according to the
height measurements and linear fittings. This is in line with the
suggestion that the eruption is driven by the HC structure.

On the other hand, the accompanying radio burst was recorded by
the Artemis radio spectrograph (Kontogeorgos et al. 2006) and
imaged by the NRH. From the spectra shown in Figure 3a, we see a
wide-band continuum emission from $\sim$ 200 to 20 MHz (10:38 to
11:00 UT). The spectra present a trend of downward drift with
time. A data gap exists from 110 $\--$ 90 MHz. It looks like that
the radio burst continues after the gap till the frequency reaches
the lower spectral end ($\sim$ 20-30 MHz).

The high frequency counterpart of the radio burst is not
clear possibly due to the low sensitivity of the Artemis
instrument as well as the strong radio interference. Yet,
according to the much more sensitive measurements of NRH, there
exist weaker yet significant emissions from 150 to 445 MHz. From
Figure 3, we see that the maximum intensity of these emissions is
higher at lower frequencies, reaching $\sim$ 100 sfu at 445 MHz
and $\sim$ 1200 sfu at 150 MHz. The burst at 445 MHz, the highest
NRH observing frequency, starts around 10:34 UT, and lasts for
$\sim$ 15 minutes. At lower NRH frequencies, the starting and peak
times are delayed, and the duration becomes longer reaching $\sim$
25 minutes for 150 MHz. This is in general consistent with
the characteristics of a broadband emission that drifts towards
lower frequency. Therefore, we conclude that the radio burst,
starting from around 10:34 UT (at 445 MHz) and drifting to lower
frequencies, represents a t-IVm event.

The temporal evolution of the polarization levels of the t-IVm
burst is shown in Figures 3d and 3e, which are calculated as the
ratio of the sum of Stokes-V over the sum of Stokes-I within the
region defined by the contour level of 95 \% of the maximum $T_B$
in each NRH image. We see that the polarization levels vary
significantly, being left-handed at earlier stage (before $\sim$
10:42 UT), becoming right-handed and increasing gradually later,
up to values $\sim$ 10-20\% for 150 MHz and 50\% for 327 MHz
($\sim$40\% for 445 MHz) at $\sim$ 10:44 UT. The emission spectra
given by the total flux intensities at different NRH frequencies
are shown in Figure 3f during the t-IVm. The spectra are generally
power-law like with an spectral index varying from -1 to -3. The
spectral shape varies considerably with time.

The corresponding $T_B$ images can be viewed from Figure 4 and the
accompanying animation. The t-IVm sources can be identified
through all NRH channels, consistent with the broadband continuum
characteristic. The $T_B$ maximum ($T_{Bmax}$) reaches higher
values at lower frequencies, being $\sim$ 10$^{7}$ , 10$^{8}$, and
10$^{9}$ K at 432, 360, and 150 MHz, respectively. These high
$T_B$ values are important to further infer the emitting
mechanism. From 10:39 to 10:43 UT the sources move outward (along
with the plasmoid, see below). They disappear earlier and reach
lower altitude at higher frequency. In general, lower frequency
sources have higher $T_B$, consistent with the above description
of the radio fluxes and the power-law like spectra. The
NRH imaging data show a continuous motion and transition of radio
sources from high to low frequencies, supporting the above
conclusion that the t-IVm burst contains a high-frequency (200 -
445 MHz) component.

\section{Correlation of the radio sources and the eruptive HC structure}
In this section, we further examine the spatial correlation of the
radio burst and the eruptive structure. To do this, in Figure 5 we
plot the contour levels at 95$\%$ of $T_{Bmax}$ at different
frequencies of a certain time together, and superpose them onto
the closest AIA 131~\,\AA{} images. The sources can be roughly
separated into two groups. The first group of sources, at higher
NRH frequencies (445 - 360 MHz), line up together on the northern
side of the dimming HC foot region without considerable motion.
This group does not belong to the t-IVm burst of study.
The other group of sources, observable at almost all NRH
frequencies, first move outward as a compact source without
significant spatial dispersion and then get spatially dispersed.
These sources correspond to the t-IVm burst of study.

According to the dynamical evolution and spatial correlation with
the eruptive structure of the t-IVm related radio sources, we can
separate the t-IVm event into two stages. In the first stage (see
panels of Figure 5a, before $\sim$ 10:44 UT), the sources at
different frequencies are largely co-spatial with the plasmoid,
i.e., the top part of the HC, without obvious spatial dispersion
as mentioned. In the second stage (Figure 5b, from $\sim$ 10:44 to
10:47 UT), the sources start to extend sunwards along the curved
side arcade of the HC with considerable spatial dispersion. From
the HC top part to its side arcade, the sources line up from high
to low frequencies. For example, the 270 MHz source is still
located at the top part while the lower frequencies shift sunward
(in the projection plane). The projected distance between the 270
MHz and 150 MHz source centroids is about 120 arcsecs at 10:45 UT.

The height measurements of the 228 MHz sources along S1 have been
over-plotted in Figure 2b with red vertical lines. The
comparison with the plasmoid locations indicate the spatial
coincidence between the HC plasmoid and the radio sources before
10:44 UT. Later, the sources get spatially dispersed and the HC
structure becomes very faint and almost invisible. To reveal their
spatial correlation, in the middle (10:45 UT) and right (10:46 UT)
panels of Figure 5b we overlay the outline of the side arcade of
the HC observed earlier at 10:40 UT (see Figure 5a). Assuming a
propagation speed of 300 km s$^{-1}$, the eruptive structure can
propagate a distance of 125 (150) arcsecs in 5 (6) minutes. It can
be seen that the radio sources are consistent with the expected
location and shape of the outward-propagating side arcade of the
HC.

These observations indicate that the t-IVm sources are
emitted by energetic electrons that are first trapped within the
plasmoid structure and then get scattered along the expanding
side-arcade structure of the eruptive HC.

Around 10:46 UT, the sources (360-270 MHz) start to appear
at the southeastern side of the AR region. Later, more sources are
present there and yield a nice formation, with lower frequency
sources located at higher altitudes. The morphology of the
formation is analogous to the leg of an open or a large-scale
closed loop, pointing towards the AR or the dimming foot (see
Figure 5c). For these radio sources, it is not possible to
determine their spatial correlation with the eruptive HC structure
which already becomes completely invisible in AIA images. This
part of the burst should not be taken as a part of the t-IVm burst
that shifts towards lower frequency. The radio emitting electrons,
however, might be physically related. For instance, the t-IVm
emitting electrons may get dumped to the leg of the HC structure
and continue to emit there when the confinement at the top and the
flank fails. It is also possible that the radio burst is a part of
the following decimetric flare continuum emission (see Figure 3a),
and the energetic electrons are not related to those accounting
for the t-IVm burst. Available data sets do not allow us to reach
a conclusive statement.

\section{Discussion}
With the above analysis, we present strong evidence of the
HC being the t-IVm emitting structure. According to the GBM data
presented in Figure 2a at 27.3$-$50.9 keV, the HXR-emitting
nonthermal electrons are present from 10:30 to 10:50 UT. During
this interval, the plasmoid forms and erupts, likely a result of
the flare reconnection, which may contain the most twisted
magnetic field lines within the HC structure. This makes it an
efficient trap of flare-injected energetic electrons, as inferred
from the t-IVm sources being spatially correlated with the
plasmoid at the earlier stage. Along with the upward eruption, the
HC plasmoid may get untwisted to gradually release the
confinement. This allows the electrons to spread over a larger
section of the structure along its side arcade, as has been
manifested by the spatially-dispersed radio sources in FIgure 5b.

Note that the CME structures have been imaged at the radio
wavelength in earlier studies. Bastian et al. (2001) proposed the
radio CME concept by investigating one specific event, in which
the radio sources present an expanding loop shape. Later studies
(Maia et al. 2007; D$\acute{e}$moulin et al. 2012) present another
event with similar characteristics, and claim that the radio
sources correspond to the CME cavity. The sources in these two
events are co-spatial at different frequencies, with a large
spatial extension (several solar radii), a relatively low $T_B$
($\le$ 10$^6$ K), and a power-law spectra of radio fluxes. Based
on these observations, they suggest that the bursts are given by
the incoherent synchrotron emission mechanism. Similar
observational characteristics are obtained for the latest imaging
studies on a t-IVm burst combining the NRH and AIA data (Tun {\&}
Vourlidas 2013; Bain et al. 2014), as introduced earlier. They
conclude that the t-IVm burst is given by the gyro-synchrotron
emission.

Here we find very different characteristics. First, the radio
sources at different frequencies are initially co-spatial with
each other, yet later they spread across the emitting structure
and get spatially dispersed with an organized pattern. Each source
occupies a relatively small part, and together they delineate a
larger section of the structure. The $T_B$ is much higher than
reported by those studies (consistent with some other earlier
observations on the t-IVm radio bursts (Stewart et al. 1978;
Duncan et al. 1980)). Therefore, we suggest that our event points
to a different emitting mechanism, possibly a coherent one to
account for the high $T_B$. Since the radio bursts here are
constrained to certain specific part of the HC, they should be
attributed to energetic electrons trapped therein. This means that
the classical plasma emission mechanism induced by fast electron
beams propagating in background plasmas may not play a role here.

Another coherent emission mechanism that is associated with
trapped electrons is the electron maser cyclotron emission,
developed when electrons are lost from the trap to form a positive
gradient of the velocity distribution function along the
perpendicular direction in the velocity phase space (Wu et al.
1979). Earlier studies have used this mechanism to explain
continuum radio bursts such as the t-IV burst (Winglee {\&} Dulk
1986; Lakhina {\&} Buti 1985). Yet, further studies are required
to check whether this mechanism can properly account for the radio
characteristics reported here.

Our study presents that the t-IVm sources are carried by the
plasmoid-HC structure. Such structure is of high temperature, and
not observable with earlier solar imaging instruments. It is
therefore not known whether such structure is present in those
earlier events.

The polarization data in our event present a change of handed
sense. Since the polarization levels are strongly affected by the
angle between the magnetic field and the wave vector, we
tentatively attribute this observation to the presumed untwisting
motion of the HC, which likely results in some rotation of the
magnetic structure.

The last point of discussion is relevant to the
ionospheric effect on source positions. Firstly, the significance
of this effect declines with increasing frequency (Wild 1959;
Bougeret 1981). At frequencies above 150 MHz the effect is
generally small. Secondly, from the NRH movie accompanying Figure
4, the positional fluctuation of radio sources is not important in
comparison to the overall motion of the sources, indicating that
the effect of ionospheric irregularities is relatively weak.
Lastly, in this event the radio sources, with an overall outward
propagation, first concentrate around the plasmoid and then get
scattered gradually along the HC side structure. This pattern is
consistent with the eruption of the HC and not likely caused by
the ionospheric effect, which tends to result in systematic
position shift of sources at different frequencies. We therefore
suggest that the ionospheric effect does not affect the result of
this study.

\section{Summary} This study demonstrates that the
AIA-observed eruptive HC structure can be imaged with the NRH at
various metric wavelengths in the form of a t-IVm burst. The radio
sources at different frequencies are found to be co-spatial with
the moving HC structure. Initially, the sources are constrained at
the plasmoid which is at the top front of the HC. Then, the
sources at different frequencies spread across the structure to a
larger region and line up across the HC side arcade. Other
characteristics include a relatively-high brightness temperature
($\sim$ 10$^{7}$ $\--$ 10$^{9}$ K), a moderate polarization with a change of the handed sense, and a power-law
like spectral shape that evolves considerably with time.
Preliminary discussion on the possible emitting mechanism has been
presented, and more studies are required to further explore the
diagnostic potential of this kind of events to infer the critical
parameters/conditions of the eruptive structure.

\acknowledgements We thank the Artemis, NRH, STEREO, SDO, and SOHO
teams for providing the data. We also thank Dr. Koval and Dr.
Hillaris for their help in plotting the dynamic spectrum. This
work was supported by grants NNSFC 41274175, 41331068, and
U1431103, and NSBRSF 2012CB825601.

\begin{figure}[t]
\includegraphics[trim = 22mm 130mm 10mm 2mm, clip]{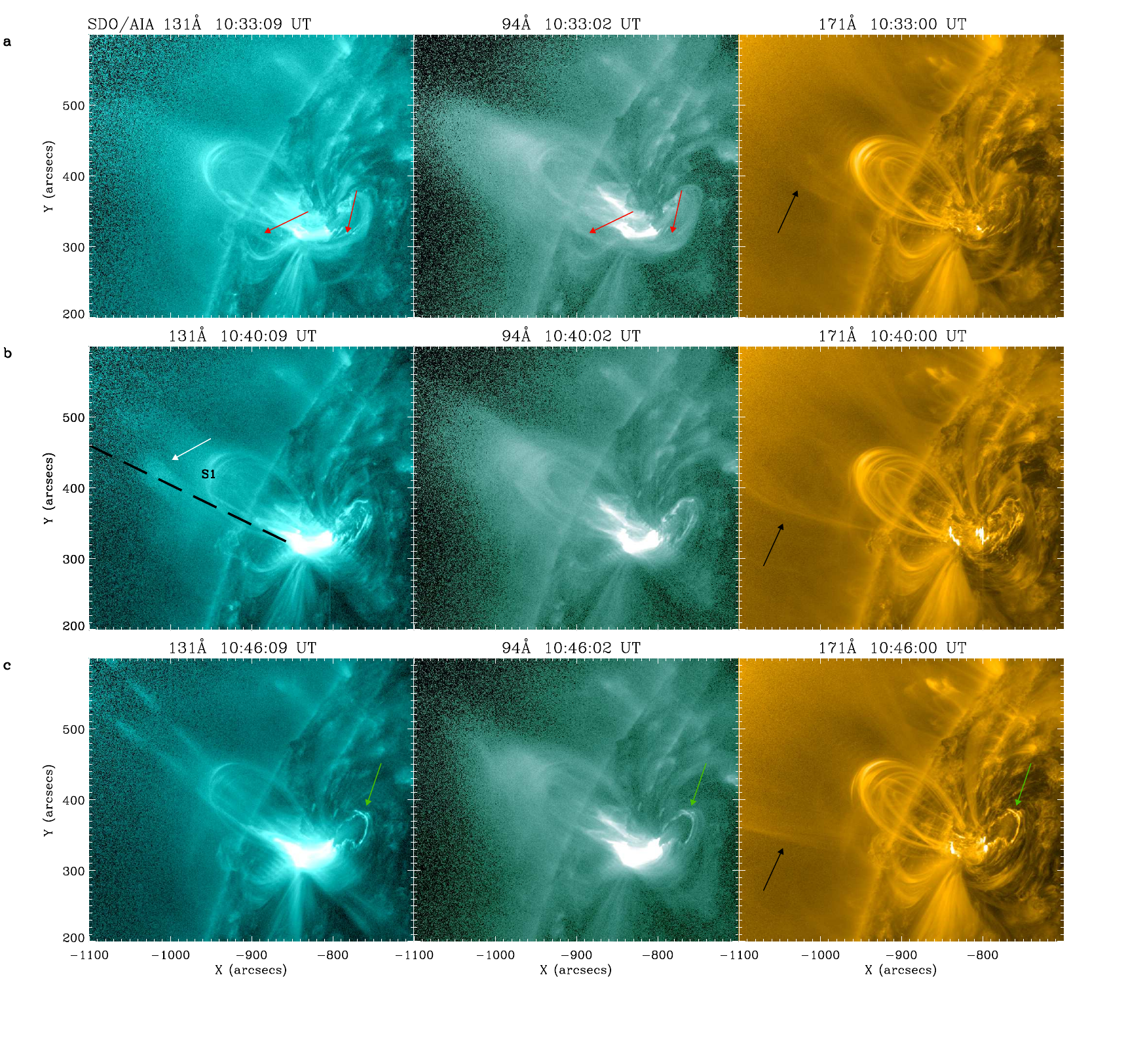}
\caption{The HC eruption recorded by AIA at 131, 94, and
171~\,\AA{} at (a) 10:33 UT (to show the slowly rising HC, red
arrows), (b) 10:40 UT (to show the eruptive plasmoid-HC structure,
white arrows), and (c) 10:46 UT (to show the dimming foot region,
green arrows). S1 is for the distance maps shown in Figure 2. An
accompanying movie of both direct and base difference images is
available online.}\label{Figure 1}
\end{figure}

\begin{figure}
\includegraphics[trim = 7mm 50mm 5mm 100mm, clip]{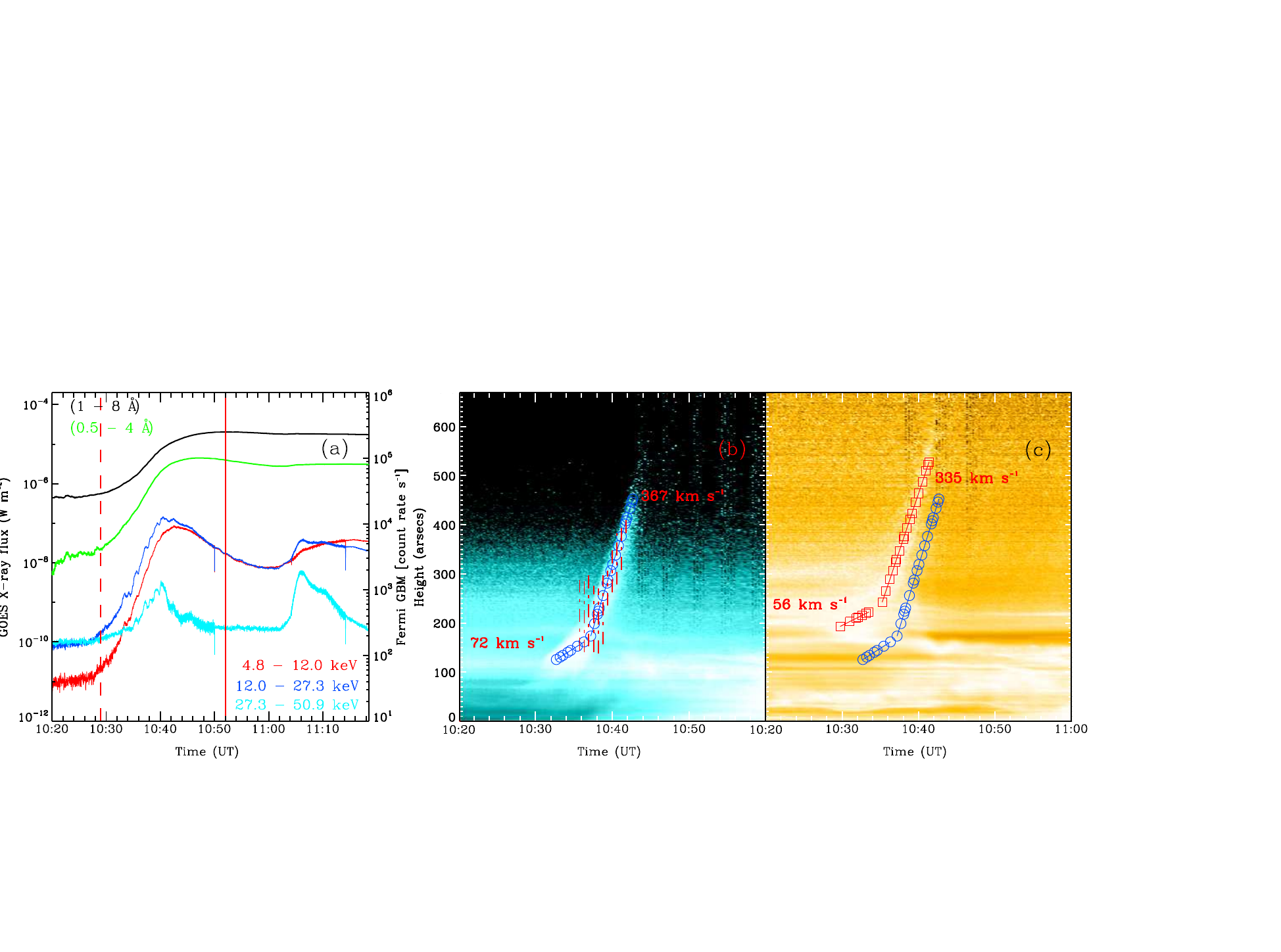}
\caption{(a) The GOES SXR light curves (1-8~\,\AA{}, and
0.5-4~\,\AA{}) and the Fermi GBM data (detector n2) at
4.8-12.0, 12.0-27.3, and 27.3-50.9 keV. The GBM data have been
averaged at a time resolution of 2 s. The direction angle of the
n2 detector to the Sun is stable and around 60 degrees during the
time of interest (10:30-11:00 UT). The two vertical lines
indicate the flare start (10:29) and peak (10:52) time. (b-c)
Distance-time maps along the slice S1 observed at AIA 131 and
171~\,\AA{}. The velocities are given by linear fitting. Red
vertical lines in panel (b) represent the NRH source positions
given by the 95{\%} $T_{Bmax}$ contour levels at 228
MHz.}\label{Figure 2}
\end{figure}

\begin{figure}
% \epsscale{1.}
\includegraphics[trim = 2mm 10mm 15mm 120mm, clip]{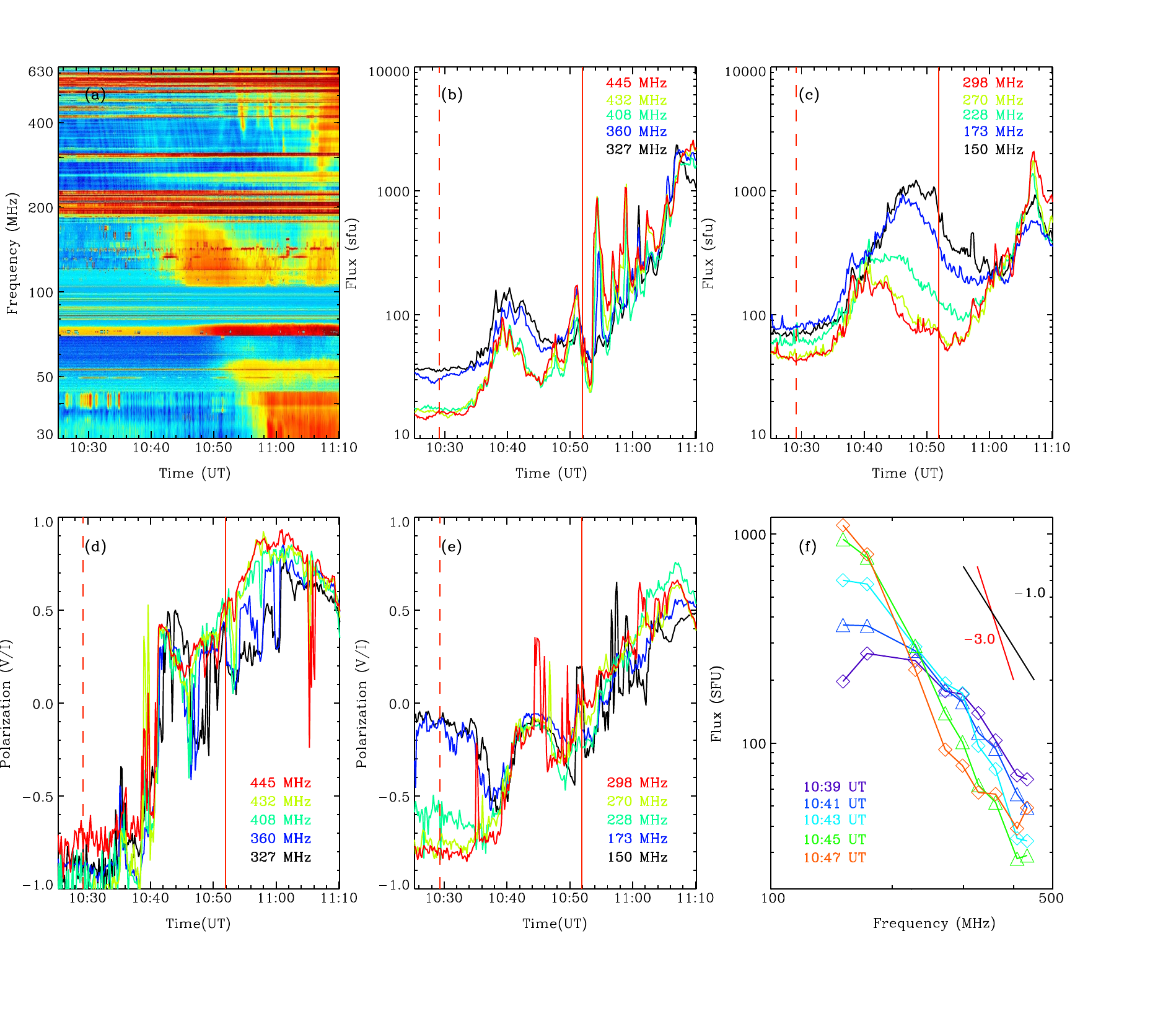}
\caption{Characteristics of the t-IVm burst as recorded by Artemis
and NRH. (a) the Artemis dynamic spectrum; (b-c) total NRH fluxes
and (d-e) polarization levels; (f) the emission spectra at
selected moments. The two vertical lines indicate the flare start
and peak time.}\label{Figure3}
\end{figure}

\begin{figure}
\includegraphics[trim = 5mm 10mm 5mm 120mm, clip]{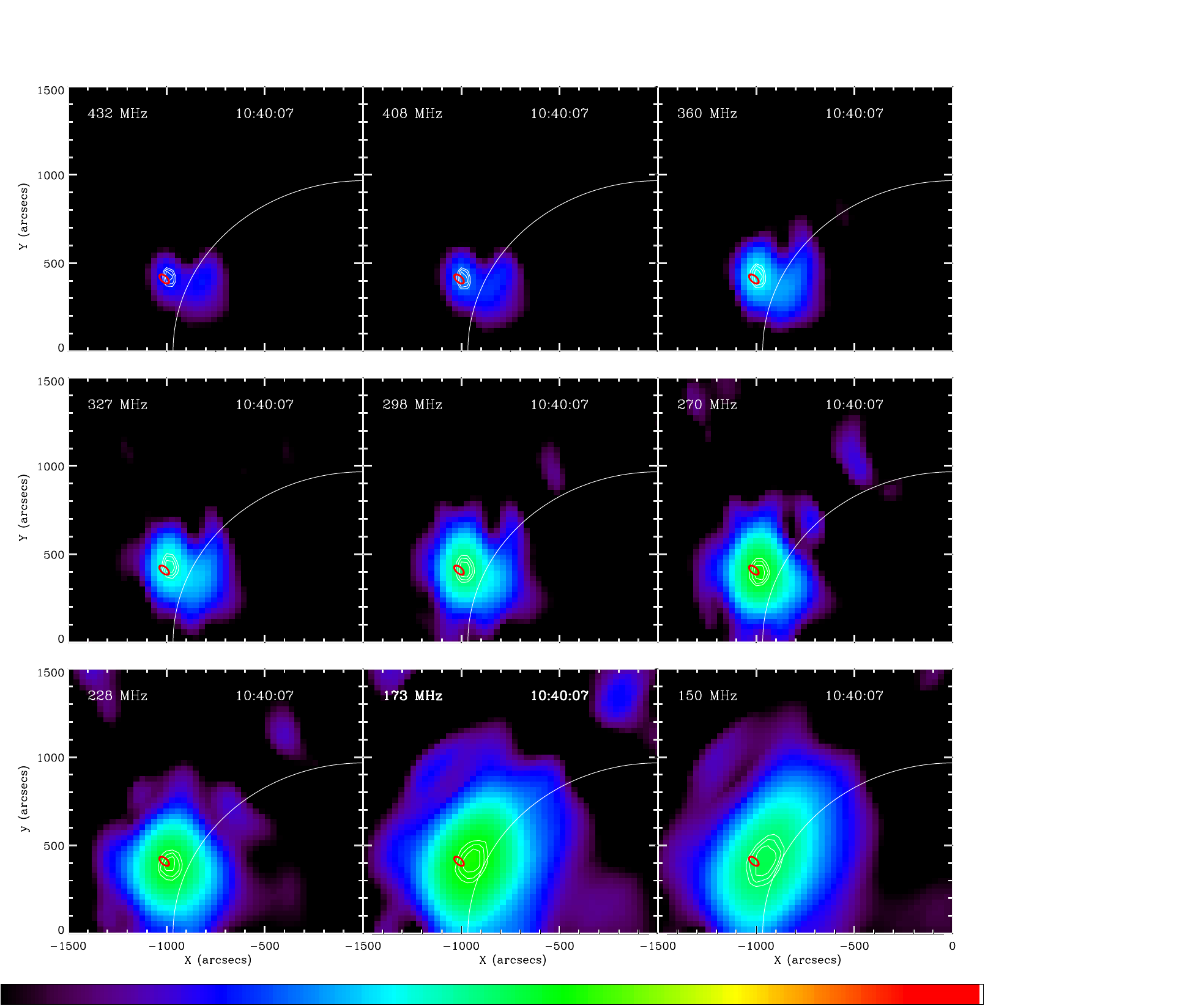}
\caption{NRH images at different frequencies at 10:40 UT. The
contour levels represent 85 \%, 90 \%, and 95 \% of $T_{Bmax}$.
The location of the HC-plasmoid observed by AIA at the same time
is denoted with a red ellipse. An accompanying movie is available
online.}\label{Figure 4}
\end{figure}

\begin{figure}
\includegraphics[trim = 5mm 10mm 5mm 120mm, clip]{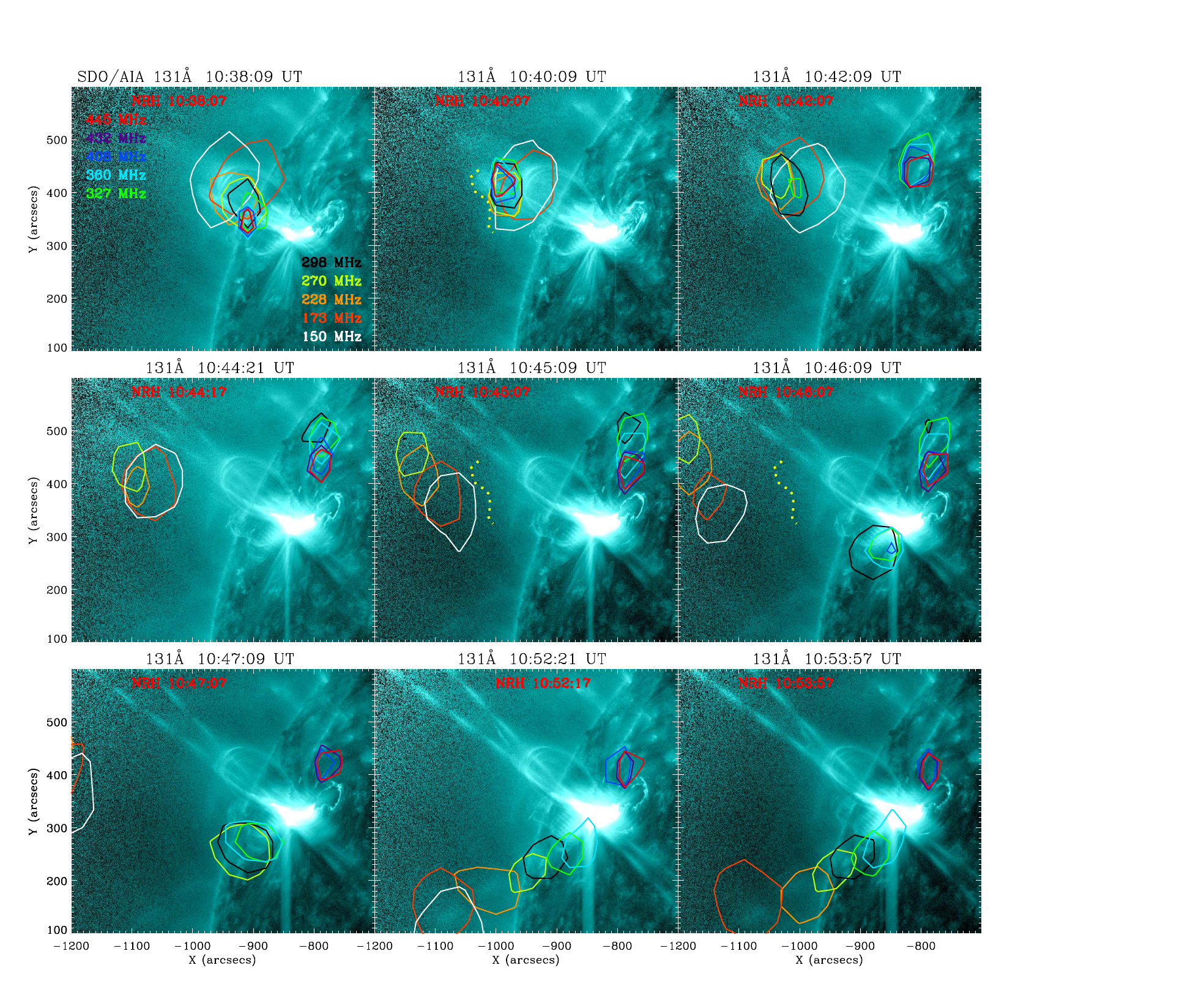}
\caption{(a-c) Temporal evolution of the NRH sources, superposed
onto the closest-in-time AIA 131~\,\AA{} images. The NRH data are
represented by the 95$\%$ $T_{Bmax}$ contours. The yellow
dotted lines present the outline of the side arcade inferred at
10:40 UT. An accompanying movie (with AIA difference images) is
available online.}\label{Figure 5}
\end{figure}


\begin{thebibliography}{}
\bibitem[Bain et al.(2014)]{bain14}Bain, H. M., Krucker, S., Saint-Hilaire, P., et al. 2014, ApJ, 782, 43
\bibitem[Bastian et al.(2001)]{bastian01}Bastian, T. S., Pick, M., et al. 2001, ApJ, 558, L65
\bibitem[Bastain and Gray.(1997)]{bastain and gray97}Bastain, T. S., and Gray, D. E. 1997, JGR, 102, 14031
%\bibitem[Benz.(2002)]{benz02}Benz, A. O. 2002, Plasma Astrophysics: Kinetic Processes in Solar and Stellar Coronae, Second Edition, Kluwer Academic Publishers, Newyork
\bibitem[bougeret1981]{bougeret1981}Bougeret, J. L. 1981, A\&A, 96, 259
\bibitem[Boischot.(1957)]{boischot57}Boischot, A. 1957, Compt. Rend. Acad. Sci. Prais, 244, 1326
\bibitem[Brueckner et al.(1995)]{brueckner95}Brueckner, G. E., Howard, R. A., Koomen, M. J., et al. 1995,SoPh,162, 357
\bibitem[Chen et al.(2014)]{chen14}Chen Y., Du G. H., Feng L. et al 2014, ApJ, 787, 59
\bibitem[Cheng et al.(2011)]{cheng11}Cheng, X., Zhang, J., Liu, Y., et al. 2011, ApJ, 732, L25
%\bibitem[Cheng et al.(2013)]{cheng13}Cheng, X., Zhang, J., Ding, M. D., et al. 2013, ApJ, 763, 43
%\bibitem[Chintzoglou et al.(2015)]{chintzoglou15}Chintzoglou, G., Patsourakos, S., and Vourlidas, A. 2015, ApJ, 809,34
\bibitem[Colaninno and Vourlidas.(2015)]{colaninno and vourlidas15}Colaninno R. C. and Vourlidas A. 2015, ApJ, 815, 70
\bibitem[Demoulin et al.(2012)]{demoulin12}D$\acute{e}$moulin, P., Vourlidas, A., Pick, M., et al. 2012, ApJ, 750, 147
\bibitem[Du et al.(2015)]{du15}Du G. H., Kong X. L., Chen Y. et al 2015, ApJ, 812, 52
\bibitem[Duncan.(1980)]{duncan80}Duncan, R. A., Stewart, R. T., Nelson, G. J. 1980, in IAU symp. 91, 381
\bibitem[Duncan.(1981)]{duncan81}Duncan, R. A. 1981, SoPh, 73, 191
%\bibitem[Elmhamdi et al.(2013)]{elmhamdi13}Elmhamdi, A., Kordi, A.S., Al-Trabulsy El-Nawawy, M., et al.: 2013, New Astron. 23, 73
\bibitem[Feng et al.(2013)]{feng13}Feng S. W., Chen Y., Kong X. L. et al 2013, ApJ, 767, 29
\bibitem[Feng et al.(2015)]{feng15}Feng S. W., Du G. H., Chen Y. et al 2015, SoPh, 290, 1195
%\bibitem[Forbes.(2000)]{forbes2000}Forbes, T. G. 2000, JGR, 105, 23153
\bibitem[Harker and Pevtsov.(2013)]{harker and pevtsov13}Harker B. J. and Pevtsov A. A. 2013, ApJ, 778, 175
\bibitem[Kong et al.(2012)]{kong12}Kong X. L., Chen Y., Li G. et al 2012, ApJ, 750, 158
\bibitem[Kontogeorgos.(2006)]{kontogeorgos06}Kontogeorgos, A., Tsitsipis, P., Caroubalos, C., et al. 2006, ExA, 21, 41
\bibitem[Kuijpers.(1975)]{kuijpers75}Kuijpers, J.: 1975, Solar Phys. 44, 173
\bibitem[Lakhina and Buti.(1985)]{lakhina and buti85}Lakhina, G. S. and Buti, B. 1985, SoPh, 99, 277
\bibitem[Lemen et al.(2012)]{lemen12}Lemen, J. R., Title, A. M., Akin, D. J., et al. 2012, SoPh, 275, 17
\bibitem[Liu et al.(2014)]{liu14}Liu Y. T., Richardson J. D., Wang C., et al. 2014, ApJL, 788, L28
%\bibitem[Low.(2001)]{low2001}Low, B. C. 2001, JGR, 106, 25141
\bibitem[Low and Hundhausen.(1995)]{low and hundhausen95}Low, B. C., Hundhausen, J. R. 1995, ApJ, 443, 818
\bibitem[Maia et al.(2007)]{maia07}Maia, D. J. F., Gama, R., Mercier, C. et al. 2007, ApJ, 660, 874
\bibitem[meegan2015]{meegan2015}Meegan, C., Lichti, G., Bhat, P. N., et al. 2009, ApJ, 702, 791
\bibitem[Melrose.(1980)]{melrose80}Melrose D. B. 1980 SSRv, 26, 3
\bibitem[Nakajima et al.(1994)]{nakajima94}Nakajima, H., Nishio, M., Enome, S., et al. 1994, IEEEP, 82, 705
\bibitem[Nelson and Melrose.(1985]{nelson and melrose85}Nelson, G.J., Melrose, D.B. 1985, In: McLean, D.J., Labrum, N.R. (eds.) Solar Radiophysics, Cambridge
University Press, Cambridge, 37.
\bibitem[Parker.(1957)]{parker57}Parker, E. N. 1957, JGR, 62, 509
%\bibitem[Patsourakos et al.(2016)]{patsourakos16}Patsourakos, S., Georgoulis, M. K., Vourlidas, A., et al. 2016, ApJ, 817, 14
\bibitem[Pesnell et al.(2012)]{pesnell12}Pesnell, W. D., Thompson, B. J., Chamberlin, P. C., et al. 2012, SoPh, 275, 3
\bibitem[Priest and Frobes.(2000)]{priest and frobes00}Priest, E., Forbes, T. 2000, Cambridge University press, New york
\bibitem[Ramesh et al.(2013)]{ramesh13}Ramesh, R., Kishore, P., Mulay, S. M., et al. 2013, ApJ, 778, 30
%\bibitem[Wang et al.(2014)]{wang14}Wang, R., Liu, Y.D., Yang, Z., Hu, H. 2014, ApJ, 791, 84.
\bibitem[Smerd and Dulk.(1971)]{smerd and dulk71}Smerd, S. F. and Dulk, G. A. 1971, IAU Symp. 43, Solar Magnetic Fields, ed. R. Howard (Dordrecht: Reidel), 616
%\bibitem[Song et al.(2014)]{song14a}Song, H. Q., Zhang, J., Chen, Y., et al. 2014, ApJ, 792, L40
%\bibitem[Song et al.(2014b)]{song14b}Song, H. Q., Zhang, J., Cheng, X., et al. 2014, ApJ, 784, 48
%\bibitem[Song et al.(2015)]{song15}Song, H. Q., Chen, Y., Zhang, J., et al. 2015, ApJ, 808, L15
\bibitem[Stewart.(1978)]{stewart78}Stewart, R. T., Duncan, R. A., Suzuki, S., et al. 1978, Proc. Astron. Soc. Australia, 3, 247
\bibitem[Stewart.(1985)]{stewart85}Stewart, R. T. 1985, in Solar Radio Physics, Cambridge University press, 361
\bibitem[Sweet.(1958)]{sweet58}Sweet, P. A. 1958, IAUS, 6, 123S
\bibitem[Takano et al.(1997)]{takano97}Takano, T., Nakajima, H., Enome, S., et al. 1997, LNP, 483, 183
\bibitem[TunVourlidas.(2013)]{Tunvourlidas13}Tun, S. D., and Vourlidas, A. 2013, ApJ, 766, 130
\bibitem[Vasanth et al.(2014)]{vasanth14}Vasanth V., Umapathy S., Vršnak B. et al 2014, SoPh, 289, 251
\bibitem[Vlahos et al.(1982)]{vlahos82}Vlahos, L., Gergely, T. E., and Papadopoulos, K. 1982, ApJ, 258, 812
\bibitem[Weiss.(1963)]{weiss63}Weiss, A. A. 1963, AuJPh, 16, 526
\bibitem[wild1959]{wild59}Wild, J. P., Sheridan, K. V., \& Neylan, A. A. 1959, AuJPh, 12,
369
\bibitem[Wild and McCready.(1950)]{wild and mccready50}Wild, J.P., McCready, L.L. 1950, Aust. J. Sci. Res., Ser. A 3, 387
\bibitem[Winglee and Dulk.(1986)]{winglee and dulk86}Winglee, R. M.; Dulk, G. A. 1986, ApJ, 307, 808
\bibitem[Wu and Lee.(1979)]{wu and lee79}Wu C. S. and Lee L. C. 1979, ApJ, 230, 621
\bibitem[Wu et al.(2016)]{wu16}Wu, Z., Chen, Y., Huang, G., et al. 2016, ApJL, 820, L29
\bibitem[Zhang et al.(2012)]{zhang12}Zhang, J., Cheng, X., Ding, M. D., et al. 2012, Nat. commun., 3, 747
\end{thebibliography}
\end{document}